\documentstyle[prd,aps,twocolumn,epsfig]{revtex}

\epsfclipon

\preprint{IC preprint number}

\begin{document}

\typeout{--- Title page start ---}

\renewcommand{\thefootnote}{\fnsymbol{footnote}}

\title{Can we Measure Superflow on Quenching $^{4}He$?}

\author{E.\ Kavoussanaki and R.\ J.\ Rivers}

\address{Blackett Laboratory, Imperial College, London SW7 2BZ}

\maketitle

\renewcommand{\thefootnote}{\arabic{footnote}}

\setcounter{footnote}{0}

\typeout{--- Main Text Start ---}

\begin{abstract}

Zurek has provided a simple picture for the onset of the
$\lambda$-transition in $^{4}He$,
not currently supported by vortex density experiments. 
However, we argue that the seemingly similar argument by Zurek
that superflow in an annulus of $^{4}He$ at a quench will be measurable is
still valid.

\end{abstract}


\narrowtext

As the early universe cooled it underwent a series of phase
transitions, whose inhomogeneities have observable consequences. To
understand how such transitions occur it is necessary to go beyond
the methods of equilibrium
thermal field theory that identified the transitions in the first instance.

In practice, we often know remarkably little about the dynamics of
quantum field theories.
A simple question to ask is the following: In principle, the field
correlation length diverges at a continuous transition.  In
reality, it does not.  What happens?
Using simple causal arguments Kibble\cite{kibble1,kibble2} made
estimates of this early field ordering, because of the implications for astrophysics.

There are great
difficulties in converting predictions for the early universe into
experimental observations.
Zurek suggested\cite{zurek1} that similar arguments were applicable
to condensed matter systems for which direct experiments could be
performed.  In particular, for $^{4}He$ he argued that the
measurement of superflow at a quench provided a simple test of these ideas.
We present a brief summary of his argument.

Assume that the dynamics of the $^{4}He$ lambda-transition can be derived from an
explicitly time-dependent Landau-Ginzburg free energy of the form
\begin{equation}
F(T) = \int d^{3}x\,\,\bigg(\frac{\hbar^{2}}{2m}|\nabla\phi |^{2}
+\alpha (T)|\phi |^{2} + \frac{1}{4}\beta |\phi |^{4}\bigg),
\label{FNR}
\end{equation}
in which $\alpha (T)$ vanishes at
the critical temperature $T_{c}$.
Explicitly, let us assume the mean-field result
$\alpha (T) = \alpha_{0}\epsilon (T_{c})$, where $\epsilon = (T/T_{c}
-1)$, remains valid as $T/T_{c}$ varies with time $t$.
In particular, we first take  $\alpha (t)=\alpha (T(t))=-\alpha_{0}
t/\tau_{Q}$ in the vicinity of $T_{c}$.  Then the fundamental length and
time scales $\xi_{0}$ and $t_0$ are given from Eq.\ref{FNR} as
 $\xi_{0}^{2} = \hbar^{2}/2m\alpha_{0}$ and $\tau_{0} = \hbar
/\alpha_{0}$. It follows that
the
equilibrium correlation length $\xi_{eq} (t)=\xi_{eq} (T(t)) $ and the relaxation
time $\tau (t)$ diverge at $T_c$, which we take to be when $t$ vanishes, as
\begin{equation}
\xi_{eq} (t) = \xi_{0}\bigg|\frac{t}{\tau_{Q}}\bigg|^{-1/2},
\,\,\tau (t) = \tau_{0}\bigg|\frac{t}{\tau_{Q}}\bigg|^{-1}.
\end{equation}

Although $\xi_{eq} (t)$ diverges at $t=0$ this is not the case for
the true correlation length $\xi (t)$, which can only grow so far in
a finite time. Initially, for $t<0$, when we are far from
the transition, we can assume that the field
correlation length $\xi (t)$ tracks $\xi_{eq}(t)$ approximately.
However, as we get closer to the transition $\xi_{eq}(t)$ begins
to increase arbitrarily fast.  As a crude upper bound, the true
correlation length fails to keep up with $\xi_{eq}(t)$
by the time $-{\bar t}$ at which $\xi_{eq}$ is growing at the speed of
sound $c(t) =\xi_{eq} (t)/\tau (t)$, which determines the rate at
which the order-parameter can change.  The condition
$d\xi_{eq}
(t)/dt = c(t)$ is satisfied at $t=-{\bar t}$, where ${\bar t}
=\sqrt{\tau_{Q}\tau_{0}}$, with corresponding correlation length
\begin{equation}
{\bar\xi } =\xi_{eq}(-{\bar t}) = \xi_{0}\bigg(\frac{\tau_{Q}}{\tau_{0}}\bigg)^{1/4}.
\label{xiZ}
\end{equation}
After this time it is assumed that the
relaxation time is so long that $\xi (t)$ is essentially frozen in
at ${\bar\xi}$ until time $t\approx +{\bar t}$, when it sets the
scale for the onset of the broken phase.

A concrete realisation of how the freezing sets in is provided
by  the time-dependent
Landau-Ginzburg (TDLG) equation for $F$ of (\ref{FNR})\cite{zurek2},
\begin{equation}
\frac{1}{\Gamma}\frac{\partial\phi_{a}}{\partial t} = -\frac{\delta
F}{\delta\phi_{a}} + \eta_{a},
\label{tdlg}
\end{equation}
for $\phi = (\phi_{1} +i\phi_{2})/\sqrt{2}$,
where $\eta_{a}$ is Gaussian noise.  We can show self-consistently\cite{ray}
that, for
the relevant time-interval $-{\bar t}\leq t\leq {\bar t}$
the self-interaction term can be neglected ($\beta =0$), whereby a simple
calculation finds
$\xi\approx{\bar\xi}$ in this interval, as predicted.
It thus happens that, at the onset of the phase transition, the field fluctuations are
approximately Gaussian. The  field
phases $e^{i\theta({\bf r})}$, where $\phi ({\bf r}) =|\phi ({\bf r})| e^{i\theta({\bf r})}
$, are then correlated on the
same scale as the fields.

Consider a closed path in the bulk
superfluid with circumference $C\gg \xi (t)$.  Naively, the number of 'regions'
through which this path passes in which the phase is correlated is ${\cal N} = O(C/\xi
(t))$.  Assuming an independent choice of phase in each 'region', the
r.m.s phase difference along the path is
\begin{equation}
\Delta\theta_{C} \approx\sqrt{{\cal N}} = O(\sqrt{C/\xi (t)}).
\label{rphase}
\end{equation}

If we now consider a quench in an annular container of similar circumference
$C$ of superfluid $^{4}He$ and radius $l\ll C$, Zurek suggested that the phase locked in is
{\it also} given by
Eq.\ref{rphase}, with ${\bar \xi}$ of Eq.\ref{xiZ}.  Since the
phase gradient is directly proportional to the superflow velocity we
expect a flow after the quench with r.m.s velocity
\begin{equation}
\Delta v = O\bigg(\frac{\hbar}{m}\sqrt{\frac{1}{C{\bar\xi}}}\bigg).
\label{v1}
\end{equation}
provided $l = O({\bar\xi})$.  Although in bulk fluid this superflow will disperse, if
it is constrained to a narrow annulus it should persist, and although not large
is measurable.

In addition to this experiment, Zurek also suggested that the same
correlation length ${\bar\xi}$ should characterise the separation
of vortices in a quench.  In an earlier paper\cite{ray} one of us showed that
this is too simple.  Causality arguments are not enough, and whether
vortices form on this scale is also determined by the thermal
activation of the Ginzberg regime, in which all $^{4}He$ experiments
take place.  Experimentally, this seems to be the
case\cite{lancaster2}.

Our aim in this paper is to see whether thermal fluctuations
interfere with the prediction Eq.\ref{v1}, for which experiments
have yet to be performed.  
Again consider a circular path in the bulk fluid (in the
1-2 plane), circumference $C$, the boundary of a surface $S$.
For given field configurations $\phi_{a}({\bf x})$
the phase change $\theta_{C}$ along the path can be expressed as the surface
integral
\begin{equation}
\theta_{C} = 2\pi\int_{{\bf x}\in S} d^{2}x\,\,\rho({\bf x}),
\end{equation}
where the topological density $\rho ({\bf x})$ is given by
\begin{equation}
\rho ({\bf x}) = \delta^{2}[\phi ({\bf x})]\epsilon_{jk}\partial_{j}
\phi_{1}({\bf x}) \partial_{k}\phi_{2}({\bf x}),\,\,\,i,j=1,2
\label{rhof2}
\end{equation}
where $\epsilon_{12}=-\epsilon_{21}=1$, otherwise zero.

The ensemble average $\langle\rho ({\bf x})\rangle_{t}$ is taken to be
zero at all times $t$, guaranteed by taking
$\langle\phi_{a}({\bf x})\rangle_{t} = 0 =
\langle\phi_{a}({\bf x})\partial_{j}\phi_{b}({\bf x})\rangle_{t}$.
That is, we quench from an initial state with no rotation.
For the Gaussian fluctuations that are relevant for the times of interest\cite{ray,cal},
all correlations are given in terms of the
diagonal {\it equal-time} correlation
function $G(r, t)$, defined by
\begin{equation}
\langle \phi_{a}({\bf x})\phi_{b}({\bf 0})\rangle_{ t} =
\delta_{ab}G(r, t)\,\,\,\,\, r = |{\bf x}|.
\label{diag}
\end{equation}
The correlation length $\xi (t)$ is defined by
$G(r,t) = o(e^{-r/\xi (t)})$,
for large $r >\xi (t)$.
The TDLG does not lead to simple exponential behaviour, but there is
no difficulty in defining $\xi (t)$ in practice\cite{ray,cal}.

The variance in the phase change around $C$, $\Delta\theta_{C}$ is
determined from
\begin{equation}
(\Delta\theta_{C})^{2}
=4\pi^{2}\int_{{\bf x}\in S}d^{2}x\int_{{\bf y}\in S}d^{2}y\,\langle\rho
({\bf x})\rho ({\bf y})\rangle_{t}.
\end{equation}
The properties of densities for Gaussian fields have been studied
in detail\cite{halperin,maz}. Define $f(r,t)$ by
$f(r,t) = G(r,t)/G(0,t)$.
On using the conservation of charge
\begin{equation}
\int\,d^{2}x\,\langle\rho ({\bf x})\rho ({\bf 0})\rangle_{t} = 0
\end{equation}
it is not difficult to show, from the results of\cite{halperin,maz},
that $\Delta\theta_{C}$ satisfies
\begin{equation}
(\Delta\theta_{C})^{2}
=-\int_{{\bf x}\not\in S}d^{2}x\int_{{\bf y}\in
S}d^{2}y\,\,{\cal C}(|{\bf x}-{\bf y}|,t),
\label{per}
\end{equation}
where ${\bf x}$ and ${\bf y}$ are in the plane of $S$, and
\begin{equation}
{\cal C}(r,t) = \frac{1}{r}\frac{\partial}{\partial r}\bigg(\frac{f'^{2}(r,t)}{1 - f^{2}(r,t)}\bigg).
\label{C2}
\end{equation}

Since $G(r,t)$ is short-ranged
 ${\cal C}(r,t)$ is short-ranged also.
With ${\bf x}$ outside $S$, and ${\bf y}$ inside $S$, all the
contribution to $(\Delta\theta_{C})^{2}$ comes from the vicinity of
the boundary of $S$, rather than the whole area.
That is, if we removed all fluid except for a strip from the neighbourhood of the
contour $C$ we would still have the same result.  This supports
the assertion by Zurek that the correlation length for phase
variation in bulk fluid 
is also
appropriate for annular flow.  The purpose of the annulus 
(more exactly, a circular capillary of circumference $C$ with radius $l\ll
C$) is to stop this flow dissipating into the bulk fluid.

More precisely, suppose that
$C\gg\xi (t)$.
Then, if we take the width $2l$ of the strip around the contour to
be larger than the correlation length of ${\cal C}(r,t)$, Eq.\ref{per} can be
written as
\begin{equation}
(\Delta\theta_{C})^{2}\approx -2
\,C\int_{0}^{\infty}dr\,r^{2}\,\,{\cal C}(r,t).
\label{per2}
\end{equation}
The linear dependence on $C$ is purely a result of Gaussian
fluctuations.

Insofar as we can identify the bulk correlation with the annular correlation, instead of Eq.\ref{v1}, we have
\begin{equation}
\Delta v = \frac{\hbar}{m}\sqrt{\frac{1}{C\xi_{s}(t)}}.
\label{v3}
\end{equation}
The step length $\xi_{s}(t)$ is given by
\begin{equation}
\frac{1}{\xi_{s}(t)} = 2\int_{0}^{\infty}dr\frac{f'^{2}(r,t)}{1 - f^{2}(r,t)}.
\label{step}
\end{equation}

There are two important differences between Eq.\ref{v3} and
Eq.\ref{v1}. The first is in the choice of time for which $\Delta v$ of
Eq.\ref{v3} is to be evaluated.  In Eq.\ref{v1} the time is the time $-{\bar
t}$ of
freezing in of the field correlation. Since $\xi (t)$ does not
change much in the interval $-{\bar t}<t<{\bar t}$ we can as well
take $t = 0$. We shall argue below that for Eq.\ref{v3} a
more appropriate time is the spinodal time $t_{sp}$ at which the transition has
completed itself in the sense that the fields have begun to populate
the ground states.

Secondly, a priori there is no reason to identify $\xi_{s}(t_{sp})$
with either ${\bar\xi}$ (or even $\xi (t_{sp})$).  In particular, because ${\bar
\xi}$ in Eq.\ref{v1} is defined from the large-distance behaviour
of $G(r,t)$, and thereby on the position of the nearest singularity
of $G(k,t)$ in the $k$-plane, it does {\it not} depend on the scale
at which we observe the fluid. This is not the case for $\xi_{s}(t)$
which, from Eq.\ref{step}, explores all distance scales. Because of the fractal nature of the
short wavelength fluctuations, $\xi_{s}(t)$ will depend on how many
are included, i.e. the scale at which we look.
If we quench in an annular capillary of radius $l$ much smaller than
its circumference, we are,
essentially, coarsegraining to that scale.  That is, the observed
variance in the flux along the annulus is $\pi l^{2}\Delta v$ for
$\Delta v$ averaged on a scale $l$. We make the
approximation that that is the {\it major} effect of quenching in an
annulus. This cannot be wholly true, but it is plausible 
if the annulus is not too narrow for boundary
effects to be important.

Provisionally we introduce a coarsegraining by hand, modifying
$G(r,t)$ by damping short wavelengths $O(l)$ as
\begin{equation}
G(r,t;l) = \int d \! \! \! / ^3 k\,
e^{i{\bf k}.{\bf x}}G(k,t)\,e^{-k^{2}l^{2}}.
\label{Gl}
\end{equation}
We shall denote the value of $\xi_{s}$ obtained from Eq.\ref{Gl} as $\xi_{s}(t;l)$.
It permits an expansion in terms of the
moments of $G(k,t)\,e^{-k^{2}l^{2}}$,
\begin{equation}
G_{n}(t;l) = \int_{0}^{\infty}dk\,k^{2n}\,G(k,t)\,e^{-k^{2}l^{2}}.
\end{equation}
For small $r$ it follows that
$f'^{2}(r,t;l)/(1 - f^{2}(r,t;l))$
\begin{equation}
= \frac{G_{2}}{3G_{1}}\bigg[1 -
\bigg(\frac{3G_{3}}{20G_{2}}-\frac{G_{2}}{12G_{1}}\bigg)r^{2} + O(r^{4})\bigg].
\label{stepint}
\end{equation}

Although, for large $r$, $f'(r,t;l)^{2} = o(e^{-2r/\xi (t)})$,
we find that the bulk of the integral
Eq.\ref{step} lies in the forward peak, and that a good {\it upper} bound for
$\xi_{s}$ is given by just integrating the quadratic term,
whence
\begin{equation}
\frac{1}{\xi_{s}(t;l)}\geq\frac{1}{\xi^{min}_{s}(t;l)} =\frac{4G_{2}}
{9G_{1}}\bigg( \frac{3G_{3}}{20G_{2}}-\frac{G_{2}}{12G_{1}}\bigg)^{-1/2},
\label{xis}
\end{equation}
with the equality slightly overestimated.
In units of $\xi_{0}$ and $\tau_{0}$ we have, in the linear regime\cite{ray},
\begin{equation}
G_{n}(t;l)\approx\frac{I_{n}}{2^{n + 1/2}}\,e^{(t/{\bar t})^{2}}
\int_{0}^{\infty}dt'\,\frac{e^{-(t'-t)^{2}/{\bar t}^{2}}}{[t'
+l^{2}/2]^{n +1/2}}\frac{T(t')}{T_{c}},
\label{Gt}
\end{equation}
where 
$I_{n} = \int_{0}dk k^{2n}\, e^{-k^{2}}$.
The presence of the $T(t')/T_{c}$ term is a reminder that the
strength of the noise $\eta$ is proportional to temperature.
However, for the time scales $O({\bar t})\ll \tau_{Q}$ of interest to
us this ratio remains near to unity and we ignore it.
For small relative times the integrand gets a large
contribution from the ultraviolet {\it cutoff dependent} lower
endpoint, increasing as $n$ increases.

If we return to the Landau-Ginzberg equation Eq.\ref{tdlg} we find
that $\langle |\phi |^{2}\rangle_{t}\ll\alpha_{0}/\beta$
in the interval $ {-\bar t}\leq t\leq {\bar t}$.  Although the field has frozen in, the
fluctuations have amplitudes that are more or less uniform across
all wavelengths.  As a result, what we see depends totally on the
scale at which we look.  Specifically, from Eq.\ref{Gt}
$\xi^{min}_{s}(0;l) = O(l)$,
as shown in the lowest curve of Fig.1.

If, as suggested by
Zurek, we take  $l =O({\bar \xi})$
we recover Eq.\ref{v1} qualitatively, although a wider bore
would give a correspondingly smaller flow.
However, this is not the time at which to look for
superflow since, although the field correlation length
$\xi (t)$ may
have frozen in by $t = 0$, the symmetry breaking has not begun.

Assuming the {\it linearised}\cite{ray} Eq.\ref{tdlg} for small times $t >0$ we see that,
as the unfreezing occurs, long wavelength modes with $k^{2} < t/\tau_Q$ grow
exponentially and soon begin to dominate the
correlation functions.
How long a time we have depends on the self-coupling $\beta$ which,
through $G_{1}$, sets the shortest time scale.
This is because, at the absolute latest, $G_{1}$
must stop its exponential growth at $t = t_{sp}$, when $\langle |\phi |^{2}\rangle_{t_{sp}}$, satisfies
$\langle |\phi |^{2}\rangle_{t_{sp}} = \alpha_{0}/\beta$.
We further suppose that
the effect of the backreaction that stops the growth initially
freezes in any structure.  In Fig.1 we also show $\xi^{min}_{s}(t;l)$ for
$t= 3{\bar t}$ and $t= 4{\bar t}$, increasing as $t$ increases.

For $^{4}He$ with quenches of milliseconds the field magnitude has
grown to its equilibrium value
before the scale-dependence has stopped\cite{ray}.
For vortex formation, for which the scale is $O(\xi_{0})$,
the thickness of a vortex,
the dependence of the density on scale
makes the interpretation of observations problematic.  This is not the same here. That the
incoherent $\xi_{s}$ depends
on radius $l$ is immaterial.  The end result is that
\begin{equation}
\Delta v = \frac{\hbar}{m}\sqrt{\frac{1}{C\xi_{s}(t_{sp};l)}}.
\label{v4}
\end{equation}

We saw that the expression Eq.\ref{xis} for $\xi_{s}$ assumed that $2l$ 
is larger than 
\begin{equation}
\xi_{eff}(t;l)=\bigg( \frac{3G_{3}}{20G_{2}}-\frac{G_{2}}{12G_{1}}\bigg)^{-1/2}.
\end{equation}
Otherwise the correlations in the bulk fluid from which we
want to extract annular behaviour are of longer range than the annulus
thickness.  Numerically, we find that $\xi_{eff}(0,l)=2l$ very accurately at
$t=0$, but that  $\xi_{eff}(t,l)\geq 2l$ for all $t>0$.  
A crude way to accomodate this is to cut off the integral Eq.\ref{per2}.    
With a little effort,
we see that the effect of this is
that $\xi^{min}_{s}(t_{sp},l)$ of Eq.\ref{xis} is replaced by 
\begin{equation}
\xi^{max}_{s}(t_{sp},l) =\xi^{min}_{s}(t_{sp},l)[1-(1-4l^{2}/\xi_{eff}(t_{sp},l)^{2})^{3/2}]^{-1},
\label{ximax}
\end{equation}
greater than $\xi^{min}_{s}(t_{sp},l)$ and thereby {\it reducing} the
flow velocity for narrower annuli.  These are the dashed curves in Fig.1.
The effect is largest for small radii $l\leq {\bar\xi}$, for which the approximation
of trying to read the behaviour of annular flow from bulk behaviour
is most suspect.  A more realistic approach for such narrow
capillaries is to treat the system as one-dimensional\cite{zurek1}. 
For this reason we have only considered $l\geq
{\bar\xi}$ in Fig.1.
We would
expect, from Eq.\ref{xis}, that  $\xi_{s}(t_{sp};l)$ has an {\it
upper} bound that lies somewhere between the curves.

Once $l$ is very large, so that the power in the
fluctuations is distributed strongly across all wavelengths we
recover our earlier result, that $\xi_{s}(t_{sp};l) = O(l)$.   
In Fig.1 this corresponds to the curves becoming parallel as $l$
increases for fixed $t$.
However, the change is sufficiently slow
that annuli, significantly wider than ${\bar\xi}$, 
for which experiments are more accessible, will give almost the same
flow as narrower annuli.  This would seem to extend 
the original Zurek prediction of Eq.\ref{v1} to thicker annuli,
despite our expectations for incoherent
flow.  However, we stress again that caution is necessary, since in 
the approximation to characterise an annulus by a
coarse-grained ring without boundaries we have ignored effects in
the direction perpendicular to the annulus. In particular, 
the circular cross-section of the tube has not
been taken into account.  One consequence of this is that infinite
(non-selfintersecting) vortices in the bulk fluid have no counterpart in an annulus. 
Removing such strings will have an
effect on $\Delta\theta_{C}$, since the typical fraction of vortices
in infinite vortices is at the level of $70\%$.  However, at the
spinodal time the fluctuations in $^{4}He$ are relatively enhanced
in the long wavelengths, and such an enhancement is known to reduce
the amount of infinite vortices, perhaps to something nearer to $20\%$.  The details of this effect
(being pursued elsewhere) are
unclear but, for the sake of argument we
take the predictions of the curves in Fig.1 as a rough guide in the vicinity of 
their minima.

So far we have avoided the question as to which time curves we
should follow.  This is because $t_{sp}$ itself depends on the scale
$l$ of the spatial volume for which the field average achieves its
ground state value.  In practice variation is small, with $t_{sp}$
for $^{4}He$ varying from about $3{\bar t}$ to $4{\bar t}$ as $l$ varies from
$\xi_{0}\ll{\bar\xi}$ to $l = 10{\bar\xi}$.  Since the curves for
$\xi_{s}(t_{sp};l)$ lie so close to one
another in Fig.1 once $l\geq 4{\bar\xi}$ the scale at which the
coarse-grained field begins to occupy the ground states becomes
largely irrelevant. 

Since $\Delta v$ only depends on $\xi_{s}^{-1/2}$ it is not sensitive
to choice of $l > 2{\bar\xi}$ at the relevant $t$.
Given all
these approximations our final estimate is (in the cm/sec units of Zurek\cite{zurek1})
\begin{equation}
\Delta v\approx 0.2(\tau_{Q}[\mu s])^{-\nu /4}/\sqrt{C[cm]}
\label{vf}
\end{equation}
for radii of $2{\bar\xi} - 4{\bar\xi}$,  
$\tau_{Q}$ of the order of
milliseconds and $C$ of the order of centimetres. $\nu = 1/2$ is the mean-field critical
exponent above.  In principle $\nu$ should be renormalised to $\nu =
2/3$, but the difference to $\Delta v$ is sufficiently small that we shall
not bother.
Given the uncertainties in its
derivation the result Eq.\ref{vf} is indistinguishable  from
Zurek's\cite{zurek1} (with prefactor $0.4$),
but for
the possibility of using somewhat larger annuli.
The agreement is, ultimately, one of dimensional analysis, but the
coefficient could not have been anticipated.
How experiments can be performed, even with the wider annuli that
Eq.\ref{vf} and Fig.1 suggest, is another matter.

We thank Glykeria Karra, with whom some of this work
was done.  This work is the result of a network supported by the European
Science Foundation .

\begin{figure}
\begin{center}
\centerline*{\psfig{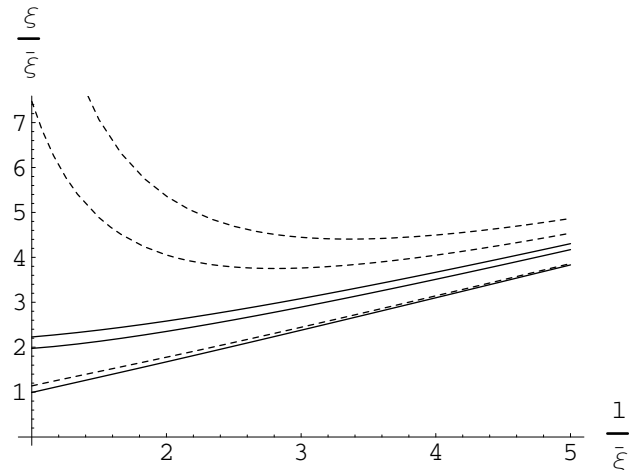}}
\end{center}
\caption{$\xi_{s}^{\min}(t,l)$ of Eq.20 (solid lines) plotted
against l, in units of $\overline{\xi}$, for t=0, 3$\overline{\xi}$ 
and 4$\overline{\xi}$.
$\xi_{s}^{\max}(t,l)$ of Eq.24 (dashed lines) plotted against l for
t=0, 3$\overline{\xi}$ and 4$\overline{\xi}$. 
In each case the higher lines correspond to higher values of time}
\end{figure}

\end{document}